# On the long-term archiving of research data

Cyril Pernet[1*], Claus Svarer[1], Ross Blair[2], John D. Van Horn[3] & Russell A. Poldrack[2]

[1] Neurobiology Research Unit, Rigshospitalet, Copenhagen, Denmark
[2] Department of Psychology & Stanford Center for Reproducible Neuroscience, Stanford University, California, USA
[3] Department of Psychology & School of Data Science, University of Virginia, Virginia, USA

\* corresponding author: wamcyril@gmail.com

**ORCID**

Cyril Pernet: 0000-0003-4010-4632
Claus Svarer: 0000-0001-7811-1825
John D. Van Horn: 0000-0003-1537-0816
Russell A. Poldrack: 0000-0001-6755-0259

**Abstract**

Accessing research data at any time is what FAIR (Findable Accessible Interoperable Reusable) data sharing aims to achieve at scale. Yet, we argue that it is not sustainable to keep accumulating and maintaining all datasets for rapid access, considering the monetary and ecological cost of maintaining repositories. Here, we address the issue of cold data storage: when to dispose of data for offline storage, how can this be done while maintaining FAIR principles and who should be responsible for cold archiving and long-term preservation.

**Introduction**

One of the goals of data curation is to ensure that data are findable and accessible to both designated users and reusers, on a day-to-day basis. The frequency of access to data is what defines their temperature, with 'hot' data being data used constantly, 'warm' data as data that needs regular access, and cold data being data with little usage. While efforts have been made to provide large-scale warm data repositories, there is little discussion about what to do with data as they become colder over time, which is also related to general recommendations on data retention policies and long-term organisational sustainability (NSTC, 2022).

Digital Data Curation includes the preservation, storage and disposal of data (Higgins, 2008). By disposal, it is usually assumed the data have not been selected for long-term curation and preservation, and that data are either transferred to a separate archive or destroyed. Here we propose to add to the 'disposal' category, the possibility of 'cold' archival storage (and thus never destroyed). Because data in cold storage are unlikely to serve users on a day-to-day basis, it naturally sits within this part of the data curation life cycle. Unlike other 'disposed' data that are destroyed or not selected for curation, 'cold data' are fully curated data. Cold archival storage is necessary to ensure long-term preservation and retention, while also reducing the unnecessary cost of maintaining them as warm data.

**When to dispose of research data?**

Usage frequency is what primarily defines data temperatures. Like molecules, the more agitation there is, the hotter the temperature. Data disposal in this context becomes about defining a frequency threshold at which one should dispose of the data, and such a threshold can be understood as a trade-off between the monetary and ecological cost of warm storage vs. the expected utility of the data. As an example, we analysed the average download counts (download byte size/dataset size) for 167 unique datasets on the OpenNeuro platform (Markiewicz, et al., 2021) deposited since January 2020. Counts were realigned from the time of deposit, with ~74% of datasets having been deposited for at least 24 months, and the smallest duration being 14 months (figure 1). From the total download counts, we can observe outliers accounting for more than 16% of the downloads and representing the top ~6% of datasets. When plotting those data over time, we can observe that for those highly accessed data, the download count does not slow down, while datasets with an average or low total download count, we observe a downward trend.

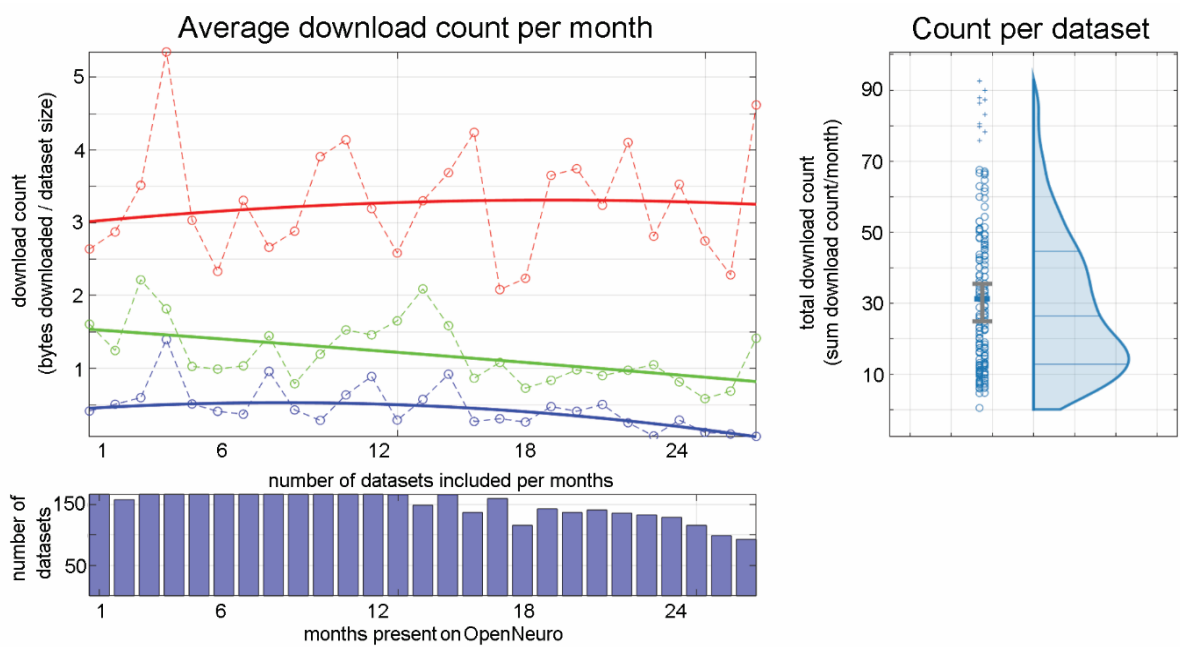

*Figure 1. Average dataset download count over time from January 2020 on https://openneuro.org/. On the left side is shown the download counts for the high access datasets (>75 total downloads, in red), those with an average access count (between 24 and 35 total downloads, in green) and those with a low access count (<18 total downloads, in blue), with thick lines representing a 2nd order polynomial fit. At the bottom, bar plots of the total number of datasets included each month. The right side shows the total download counts (circles and crosses represent individual datasets, crossed are outliers based on the median absolute deviation, the grey bars represent the mean, 31, with its 95% Highest Density Confidence Interval [24 35], and the blue area is the non-parametric kernel density estimate of the total data count).*

When it comes to decisions about which data to preserve, what matters most is their utility. Data can be downloaded often because their quality, richness and complexity allow them to be usefully further analysed or re-purposed to address new scientific questions. A good example of this is given by INDI datasets (1000 Functional Connectomes Project, ADHD-200, Autism Brain Imaging Data Exchange (ABIDE) and Consortium for Reliability and Reproducibility (CoRR) data) which have been re-analyzed many times leading to high-quality publications (913 as of March 2017), with an estimated savings of over one million dollars (Milham et al., 2018). Datasets can also be downloaded due to their simplicity, making them useful for e.g., method development or teaching, cases for which metrics are lacking. We believe those scenarios to be likely because the most downloaded datasets on OpenNeuro are not just the largest, some are also small-size datasets (as computed in figure 1, the most downloaded datasets are on average ~54GB [min 5.6Mb max 847Gb] vs ~34Gb [min 311Mb max 171Gb] for the least downloaded ones, one-sided t-test t(26.35)=0.9 p=0.18).

Usage frequency is, however, only a proxy of utility and it is self-evident that some data can be extremely valuable in very specific contexts while being rarely accessed (e.g. data on the sequence of a little-studied virus might become very important if that virus goes on to later result in a epidemic). While usage frequency can be used to decide when to dispose of data, utility is thus the concept that enables decisions about when to dispose of data and also separates data destruction from cold archiving; it cannot be approximated by usage frequency alone. Since data utility is highly field-specific and context-specific, we cannot provide general recommendations on what constitutes useful data. Given that the data discussed here regards data that have been curated and given the cost of data collection and curation, we can, however, guard against destruction and suggest cold archiving by default. Going back to the OpenNeuro dataset examples, we could consider after 24-to-36 months, moving to cold archiving the lower end of the datasets (~38%). The question that follows is whether cold data can still remain Findable, Accessible, Interoperable and Reusable (Wilkinson et al., 2016).

**Cold FAIR data**

In the context of web-based repositories, findability is highly dependent on globally unique identifiers (GUIDs), also called unique persistent identifiers (PIDs), and datasets are typically assigned such identifiers, often in the form of a Digital Object Identifier (DOI). As one moves data to cold archiving, it is essential to maintain metadata information, now indicating the data's temperature. As FAIR also focuses on machine actionability and cold archiving has been more often than not a human enterprise, a clear mechanism must be in place to request such data if needed, indicating how to request and what delays to expect. For cloud serviced repository, this can be as simply as changing data cloud tiering to cold, with the consequence of having a more expensive and longer retrieval time if access is required, but an overall lower ecological and financial cost since those data are rarely accessed.

As one thinks about long-term data preservation, we have to keep in mind that, in general, neither digital media nor institutions are reliably durable and it is thus necessary to plan for data longevity, particularly with regard to how the data will be preserved if the repository were to cease operations. This is exemplified by the fMRI Data Center project (Van Horn & Gazzaniga, 2013) which curated and archived data from over 100 complete fMRI studies between 2000 and 2006 but had to cease its activities due to the sunsetting of the NIH Human Brain Project (HBP) initiative and the subsequent lack of new funding opportunities to support continued operations. Current repositories must learn from this and implement solutions. Compared to 2006, there are two major differences. First, there exist multiple international repositories for data sharing within the neuroimaging field while fMRIDC was the only repository of its kind. Second, most neuroimaging repositories, if not all, now rely on versions of the same dataset organisation, BIDS (Brain Imaging Data Structure; Gorgolweski et al., 2016) alleviating the need for repository-centric curation models. These two elements make it easier to facilitate the transfer of datasets between repositories with little management cost, providing licencing or data usage agreements are compliant with the repository policy. It would, thus, be wise to have some coordination between repositories allowing moving datasets between them. Just as important as being able to transfer data, new tools must be developed to translate metadata schemas between repositories allowing all datasets to be findable on any repository, while pointing to where they are accessible, creating a redundant and federated network of datasets' metadata. As a repository is closing, not all data would need to be moved, only the hottest one, while others could be cold archived. Cold data can be returned to the data creators, or stored somewhere else depending on agreements that must here be in place (i.e. repositories should have that information in their data policies) with updated metadata available through partner repositories.

Interoperability and Reusability depend primarily on the data format used and data integrity. For data formats, for hot and warm data alike, we can only recommend open and well-documented formats allowing data to be retrieved and read by anyone at any time. Reusability also depends on data integrity, which is an important aspect of any data storage system. Four sources of risk have been proposed and operationalised: storage hardware; physical environment; curating institution; global environment (Altman & Landau, 2020) and all those factors must be considered. Hot and warm data integrity is often ensured by simultaneously using multiple copies of the data, for instance having regular backup copies on physically separated servers, themself using redundant arrays of independent disks, thus protecting data from drive failure. Cold archiving, however, often relies on a single copy and thus bit-level information integrity is of greater concern. A typical recommendation for cold storage is to use magnetic tape, in particular, Linear Tape Open (LTO) as those types of tapes can contain 10-14 TByte of data for about $50, with a bit-wise error rate of 1 in $10^{19}$ and up to 30-year shelf-life. These should be stored in a secured, flood-resistant and temperature-controlled environment ensuring long-term preservation. Planning for such activity in case of repository shutdown is thus necessary, with future data access planned to be managed by institutions like university libraries, possibly in relation from services allowing cloud retrieval (as e.g. Amazon Glacier Deep Archive, Google or Azure storage archive) thus allowing independent storage and retrieval when needed.

**Conclusion**

Researchers happily take responsibility for data creation and increasingly they are taking up the challenging but necessary task of data sharing. As data sharing becomes the norm, it is essential to clarify responsibilities about who ensures access to the data and in what form. Repositories should clearly state for how long data are guaranteed or just likely to be shared, state who decides when data should move to cold storage (using which criteria) and who is responsible for cold storage (Currie & Kilbride, 2021). Once those have been decided, simple steps can be made to keep data FAIR: i) keep metadata alive by making them available through multiple sources with the PID ii) have mechanisms in place to retrieve cold data and (iii) have procedures ensuring cold data physical integrity.

**Acknowledgements**

C.R.P. is supported by the Novo Nordisk Fonden NNF20OC0063277.